\documentclass{aa}  

\usepackage{epsfig}
\usepackage{graphicx}
\usepackage{natbib}

\bibpunct{(}{)}{;}{a}{}{,} % to follow the A&A style
\usepackage{txfonts}
\begin{document}

   \title{Chromospheric evaporation flows and density changes deduced from Hinode/EIS during an M1.6 flare}

   \author{P. G\"{o}m\"{o}ry\inst{1},
           A. M. Veronig\inst{2},  
           Y. Su\inst{2,3},
           M. Temmer\inst{2},
           \and
           J. K. Thalmann\inst{2}
          }
   \authorrunning{P. G\"{o}m\"{o}ry et al.}

   \institute{Astronomical Institute of the Slovak Academy of Sciences,
              05960 Tatransk\'{a} Lomnica, Slovakia\\
              \email{gomory@astro.sk}
         \and
              IGAM-Kanzelh\"{o}he Observatory, Institute of Physics, 
              University of Graz, Universit\"{a}tsplatz 5, 8010 Graz, Austria 
         \and
              Key Laboratory of Dark Matter \& Space Astronomy, Purple Mountain 
              Observatory, Chinese Academy of Sciences, 2 West Beijing Road, 
              210008 Nanjing, China
             }

   \date{Received MMMM DD, YYYY; accepted MMMM DD, YYYY}

% \abstract{}{}{}{}{} 
% 5 {} token are mandatory
 
  \abstract
   {}
   {We study the response of the solar atmosphere during a \textit{GOES} M1.6 flare
    using spectroscopic and imaging observations. In particular, we examine the evolution 
    of the mass flows and electron density together with the energy input derived from
    hard X-ray (HXR) in the context of chromospheric evaporation.}
   {We analyzed high-cadence sit-and-stare observations acquired with the \textit{Hinode}/EIS 
    spectrometer in the \ion{Fe}{xiii} 202.044\,\AA~(log\,$T$\,=\,6.2) and 
    \ion{Fe}{xvi} 262.980\,\AA~(log\,$T$\,=\,6.4) spectral lines to derive temporal 
    variations of the line intensity, Doppler shifts, and electron density during the flare. 
    We combined these data with HXR measurements acquired with \textit{RHESSI} 
    to derive the energy input to the lower atmosphere by flare-accelerated electrons.}
   {During the flare impulsive phase, we observe no significant flows in the cooler 
    \ion{Fe}{xiii} line but strong upflows, up to 80-150\,km\,s$^{-1}$, in the hotter \ion{Fe}{xvi} 
    line. The largest Doppler shifts observed in the \ion{Fe}{xvi} line were co-temporal with 
    the sharp intensity peak. The electron density obtained from a \ion{Fe}{xiii} line 
    pair ratio exhibited fast increase (within two minutes) from the pre-flare level of  
    5.01$\times$10$^{9}$\,cm$^{-3}$ to 3.16$\times$10$^{10}$\,cm$^{-3}$ during 
    the flare peak. The nonthermal energy flux density deposited from 
    the coronal acceleration site to the lower atmospheric layers during the flare peak was 
    found to be 1.34$\times$10$^{10}$\,erg\,s$^{-1}$\,cm$^{-2}$ for a low-energy cut-off 
    that was estimated to be 16\,keV. 
    During the decline flare phase, we found a secondary intensity and density peak of lower 
    amplitude that was preceded by upflows of $\sim$15\,km\,s$^{-1}$
that
were detected in both 
    lines. The flare was also accompanied by a filament eruption that was partly 
    captured by the EIS observations. We derived Doppler velocities of 250-300\,km\,s$^{-1}$ 
    for the upflowing filament material.}
   {The spectroscopic results for the flare peak are consistent 
    with the scenario of explosive chromospheric evaporation, although a comparatively low 
    value of the nonthermal energy flux density was determined for this phase of the flare. 
    This outcome is discussed in the context of recent hydrodynamic simulations. It provides 
    observational evidence that the response of the atmospheric plasma strongly depends on the 
    properties of the electron beams responsible for the heating, in particular the steepness 
    of the energy distribution. The secondary peak of line intensity and electron density detected 
    during the decline phase is interpreted as a signature of flare loops being filled by expanding 
    hot material that is due to chromospheric evaporation.}

   \keywords{Sun: flares --
             Sun: chromosphere --
             Sun: corona --
             Sun: UV radiation
            }

   \maketitle
%
%________________________________________________________________

\section{Introduction} \label{Introduction}

Flares are among the most energetic events on the Sun. According to 
the standard eruptive flare model  
(\citealt{1964NASSP..50..451C,    % Carmichael 1964
          1968IAUS...35..471S,    % Sturrock   1968
          1974SoPh...34..323H,    % Hirayama   1974
          1976SoPh...50...85K};   % Kopp & Pneuman 1976
see also reviews by
\citealt{2008LRSP....5....1B,     % Benz 2008
         2011SSRv..159...19F},    % Fletcher et al. 2011
and references therein), 
the energy to power solar flares is stored in the nonpotential coronal 
magnetic fields 
\citep[e.g.,][]{2000JGR...10523153F,        % Forbes 2000
              2011SSRv..158....5H,        % Hudson 2011
              2014A&ARv..22...78W}.      % Wiegelmann et al. 2014
The mechanism that releases magnetic energy is known as magnetic reconnection. 
Observational as well as theoretical aspects of this mechanism have 
been studied in detail 
\citep[e.g.,][]{1963ApJS....8..177P,        % Parker 1963
              1993A&A...271..292D,        % Demoulin et al. 1993
              1993SoPh..146..127L,        % Litvinenko & Somov 1993
              2001ApJ...546L..69Y,        % Yokoyama et al. 2001
              2006SoPh..238..347A,        % Aulanier et al. 2006
              2006A&A...446..675V,        % Veronig et al. 2006
              2013NatPh...9..489S,        % Su at al. 2013
              2014ApJ...784..144D,        % Dudik et al. 2014
              2014ApJ...788...85V,        % van Driel-Gesztelyi et al. 2014     
              2015ApJ...804L...8L}.       % Li and Zhang 2015
 
The vast amount of energy released during magnetic reconnetion in the 
corona is converted into heating of the surrounding plasma and accelerating 
particles to nonthermal energies. The flare-accelerated particles are 
guided by the ambient magnetic field and progress downward to the lower 
atmosphere, encountering a denser environment. They are effectively stopped 
at chromospheric heights where their energy is dissipated by Coulomb 
collisions with the ambient thermal particles    
\citep{1971SoPh...18..489B,                % Brown et al. 1971 
       1976SoPh...50..153L}.               % Lin & Hudson 1976
This rapid energy deposition causes the chromospheric plasma
to be  
intensely heated to coronal and flare (10$^{7}$\,K) temperatures, expanding 
upward, and thus to fill the coronal loops. This process is called chromospheric 
evaporation and was first proposed by 
\cite{1968ApJ...153L..59N}                 % Neupert 1968 
to explain the observed correlation between the thermal and integrated 
nonthermal flare emission 
\citep[e.g.,][]{2002A&A...392..699V}.       % Veronig et al. 2002   

The process of chromospheric evaporation was investigated in a number of theoretical 
studies. Hydrodynamic simulations revealed two evaporation regimes that are separated by an 
energy flux density threshold at roughly 10$^{10}$ erg\,cm$^{-2}$\,s$^{-1}$
\citep[e.g.,][]{1985ApJ...289..414F}.        % Fisher er al. 1985
So-called gentle evaporation occurs if 
the energy that is deposited in the chromosphere by nonthermal electrons is lower 
than this threshold. The heated chromospheric material then slowly 
expands upward with velocities of several tens of km\,s$^{-1}$. In addition, gentle 
evaporation may be also a result of thermal conduction from the hot flaring corona. 
Signatures of gentle evaporation were observed mainly during weaker flares (i.e., \textit{GOES} 
C-class and weaker) as well as during the pre-flare and late phase of stronger events 
\citep{1999ApJ...521L..75C,                % Czaykowska et al. 1999
       2004ApJ...613..580B,                % Brosius & Phillips 2004 
       2006ApJ...642L.169M}.               % Milligan et al. 2006 ApJL 642, 169 

If the energy flux densities deposited in the chromosphere exceed 10$^{10}$ erg\,cm$^{-2}$\,s$^{-1}$, 
explosive evaporation takes place. In this case, the radiative cooling is 
insufficient and the plasma is rapidly heated to coronal and flare (10$^{7}$\,K) 
temperatures. As a 
consequence, the local gas pressure rises significantly, yielding an explosive 
upward expansion of the chromospheric plasma at velocities reaching several hundred 
km\,s$^{-1}$. To regain momentum balance with the hot plasma upflows, 
the cooler material of the underlying layers is pushed downward. 
Simulations predict that downflows are noticeable only in spectral 
lines formed below transition region temperatures and that they reach 
velocities of several tens of km\,s$^{-1}$ because of the higher mass and 
inertia of the chromospheric material 
\citep{1985ApJ...289..414F,                % Fisher er al. 1985
       2005ApJ...630..573A,                % Allred et al. 2005
       2015ApJ...808..177R}.               % Reep et al. (2015) 
Observational evidence for the proposed momentum balance during explosive evaporation 
was found for several events 
\citep{1988ApJ...324..582Z,                % Zarro et al. 1988
       2006A&A...455.1123T,                % Teriaca et al. 2006, 
       2006ApJ...638L.117M}.               % Milligan et al. 2006 ApJL 638, 117
Other observational studies, however, also revealed downflows at coronal temperatures 
\citep{2009ApJ...699..968M,                % Milligan & Dennis 2009 
       2013ApJ...766..127Y},               % Young et al. 2013 
suggesting that the flaring atmosphere is very dynamic and complex on small spatial scales 
\citep[e.g.,][]{2010ApJ...719..655V}.       % Veronig et al 2010 
Simplified hydrodynamic simulations therefore probably fall short of adequately 
describing all responses of the flaring atmosphere.

In general, a characteristic signature of chromospheric evaporation are  observed spectral blueshifts during the impulsive phase of flares as 
the consequence of heated plasma rising into the corona. This has been observed more than 30 years ago 
\citep{1980ApJ...239..725D,                % Doschek et al. 1980 
       1980ApJ...241.1175F,                % Feldman et al. 1980 
       1982SoPh...78..107A}.               % Antonucci et al. 1982 
However, one-dimensional simulations of chromospheric evaporation in a single flare 
loop predict a blueshift of the spectral line profiles as a whole 
during the early phases of flares 
\citep{1987SoPh..110..295E,                % Emslie & Alexander 1987
       1989SoPh..123..161M}.               % McClements & Alexander 1989       
This contradicts the majority of the earliest observational findings, which only
revealed an asymmetry in the blue wing of the spectral line profiles (commonly 
interpreted as evidence of upflows of several hundred km\,s$^{-1}$), together with 
a dominant static spectral component
\citep[e.g.,][]{1999mfs..conf..331A}.       % Antonucci et al. 1999    
This obvious discrepancy is solved when a multi-thread, fine-structured flare loop 
model is considered 
\citep{1997ApJ...489..426H,                % Hori et al. 1997 
       2005ApJ...629.1150D,                % Doschek & Warren 2005
       2005ApJ...618L.157W}.               % Warren & Doschek 2005 
In this case, the asymmetric line profiles are the result of superposed spectral components at different velocities that are emitted from particular threads 
within the flare loop envelope. We stress, however, that recent 
results based on high-resolution measurements clearly demonstrate the existence 
of entirely blueshifted spectral profiles during the onset phases of flares 
\citep{2006SoPh..234...95D,                % Del Zanna et al. 2006 
       2011A&A...526A...1D,                % Del Zanna et al. 2011 
       2013ApJ...762..133B,                % Brosius 2013 
       2015ApJ...803...84P}.               % Polito et al. 2015

Another aspect of the chromospheric evaporation process is the enhancement of the 
electron density measured at high temperatures. Electron densities 
are often determined based on methods that use density-sensitive line ratios under the assumption that 
one of these lines arises from a meta-stable level
\citep[for more details see, e.g.,][]{1992str..book.....M}.   % Mariska 1992
It is important to determine them accurately to understand the heating and  
cooling of flaring plasma. Time-dependent studies of the density variations during flares 
are quite rare. One of the first results were published by 
\cite{1980ApJ...238L..43M}                 % McKenzie et al. 1980
and
\cite{1981ApJ...249..372D},                 % Doschek et al. 1981
who used data from the SOLEX instrument onboard {\it P78-1}. 
\cite{2011A&A...532A..27G}                 % Graham et al. 2011
used EIS observations to study the temporal evolution of densities during a C6.6 flare, but with 
a limited time-resolution of 150\,s. Recently, 
\cite{2012ApJ...755L..16M}                 % Milligan at al. 2012
presented techniques for determining time-dependent measurements of the electron density using
high-temperature ($\sim$12\,MK) density-sensitive line pairs measured with the \textit{SDO}/EVE
instrument. They found that the most reliable line pairs detected by the EVE full-Sun measurements 
are sensitive to densities in the range 10$^{11}$-10$^{14}$\,cm$^{-3}$ and are therefore suitable 
to study density variations during X-class flares, where such enhancements of electron densities 
may develop. However, electron densities exceeding 10$^{11}$\,cm$^{-3}$ (i.e., the low-density limit for 
the EVE measurements) are rather unrealistic for M-class flares.

In general, flare spectroscopy can be performed in two ways: raster and sit-and-stare 
observations. Most of the existing 
studies are based on rasters. This observing mode provides good insight into the spatial 
distribution of the response of the flaring atmosphere, but the data are acquired only 
with very limited temporal resolution (order of minutes) for dynamic events like flares. 
Therefore, the locations of the main energy deposition (flare kernels) are often missed 
in raster scans, as they have only short lifetimes. To study the temporal evolution 
of the plasma properties in flare kernels, sit-and-stare observations are preferable. 
Since the slit covers only a very limited and fixed field of
view, it is difficult 
to obtain data across entire flare kernels, however. Thus, flare studies based on sit-and-stare 
observations are still rare
\citep[e.g.,][]{2001ApJ...555..435B,         % Brosius 2001
               2009A&A...505..811B,         % Berkebile-Stoiser et al. 2009
               2010ApJ...719..655V,         % Veronig et al. 2010 
               2013ApJ...762..133B,         % Brosius 2013
               2015ApJ...810...45B}.        % Brosius and Daw 2015

We here present a study on the chromospheric evaporation process during an M-class 
flare using one such rare data set of sit-and-stare measurements from \textit{Hinode}/EIS, 
where the slit was placed across a flare kernel. We focus on the flare-induced Doppler velocities 
and densities acquired with high cadence. The spectroscopic findings are discussed in the context 
of high-cadence EUV imaging by AIA/\textit{SDO} and the characteristics of the flare-accelerated 
electrons deduced from \textit{RHESSI} hard X-ray images and spectra. 
%__________________________________________________________________

\section{Data and data reduction} \label{Data and data reduction}

The data-set presented here was obtained within the Hinode Observing 
Plan HOP-180\footnote{http://www.isas.jaxa.jp/home/solar/hinode$\_$op/hop.php?hop=0180}
that was performed during several days in February 2011. This observing program was 
specifically designed to study spectroscopic properties and dynamics of large-scale 
coronal waves (so-called EIT waves) propagating through the quiet corona by combining 
high-cadence spectroscopy with high-cadence multiwavelength imaging 
\citep{2011ApJ...743L..10V}.                      % Veronig et al. 2011   
The main observing target was AR 11158, which became the first major flaring region of 
solar cycle 24. The magnetic structure and evolution of this active region as 
well as the properties of the M- and X-class flares it produced have been studied 
in a number of papers  
\citep[e.g.,][]{2011SPD....42.2213Y,  % Young et al. 2011
               2011ApJ...734L..15K,  % Kosovichev 2011
               2011ApJ...738..167S,  % Schrijver et al. 2011 
               2012ApJ...757..149S,  % Sun et al. 2012
               2013ApJ...766..127Y,  % Young et al. 2013
               2013ApJ...770...79I,  % Inoue et al. 2013
               2015ApJ...801...36S,  % Sorriso-Valvo et al. 2015
               2015ApJ...803...73I,  % Inoue et al. 2015
               2015ApJ...807..124K}. % Kuroda et al. 2015 

We concentrate on the analysis of a subset of the data obtained on 16 February 2011
between 13:38 and 15:43\,UT. This subset covers the evolution of an eruptive M1.6 flare that 
was accompanied by an EIT wave. The properties and plasma diagnostics of the EIT wave  
have been presented in 
\cite{2011ApJ...737L...4H},                        % Harra et al. 2011
\cite{2011ApJ...743L..10V},                        % Veronig et al. 2011   
and
\cite{2013SoPh..288..567L}.                        % Long et al. 2013  

The spectroscopic data were acquired with the EUV Imaging Spectrometer  
\citep[EIS;][]{2007SoPh..243...19C}                % Culhane et al. 2007 
onboard the Japanese space mission \textit{Hinode}
\citep{2007SoPh..243....3K}.                       % Kosugi et al. 2007 
The longest possible slit with height 512$^{\prime\prime}$ and width 2$^{\prime\prime}$ 
(with a pixel size of 1$^{\prime\prime}$ in the y-direction) was used in the sit-and-stare 
observing mode. The exposure time was set to 45\,s plus $\sim$4\,s for read out. The 
effective time cadence is thus $\sim$49\,s.
The standard runs of HOP-180 were performed using 11 selected spectral lines 
that cover the temperature range log\,$T$\,=\,4.7\,-\,6.7, including several line pairs 
from the same ion for density diagnostics. However, several of these lines are unsuitable for the flare study. The main goal of HOP-180 was the detection of EIT waves 
in the quiet solar corona, and therefore blended lines were also selected because 
the blends are very weak under conditions of weak intensity enhancements. But they 
become significant in flaring regions. 

Multi-Gaussian fitting can in principle be used to separate
such profiles, but this method must be applied very carefully as it is susceptible to creating 
statistically acceptable but un-physical fits to the data. Therefore it is normally applied 
to single spectra where the result can be verified, but it is not very suitable to fit longer 
time-series. We therefore  concentrate here on the spectral lines with no blends and a strong signal: 
\ion{Fe}{xiii} 202.044\,\AA~(log\,$T$\,=\,6.2) and \ion{Fe}{xvi} 262.980\,\AA~(log\,$T$\,=\,6.4). 
In addition, we used the \ion{Fe}{xiii}\,196.640\,\AA~spectral line to determine 
coronal electron densities from the \ion{Fe}{xiii} line pair. The electron densities were 
estimated using the theoretical variation of the line intensity ratio with density using the 
CHIANTI database version 7.1. 
\citep{1997A&AS..125..149D,                         % Dere et al. 1997
       2013ApJ...763...86L}.                        % Landi et al. 2013
After the theoretical ratio was known, the final density maps were calculated using the EIS 
routine eisdensity.pro.

The EIS data were first corrected for photometric effects and calibrated using the eis$\_$prep.pro 
routine, which is part of the SolarSoftWare. The wavelength drift was compensated for using the 
so-called HK method described in  
\cite{2010SoPh..266..209K}.                         % Kamio et al. 2010 
Then the spectral profiles were fit by a single -Gaussian function with a linear background to obtain the amplitude of the profile, integrated 
intensities, background intensities, Doppler shifts, and spectral widths. The spatial offset 
in the solar y-direction among the data detected at different wavelengths was also compensated for. 
Because EIS provides no absolute wavelength scale, the determination of a rest wavelength is 
difficult especially in active regions. In our case, the longest slit that also covered quiet 
areas was used, and the zero reference of the Doppler shifts was calculated as the average 
value of the Doppler shifts from quiet-Sun regions. We adopted the formalism that positive 
velocities (redshifts) represent motions toward the solar surface and negative velocities 
(blueshifts) denote motions into the corona.

The imaging was performed using the Atmospheric Imaging Assembly 
\citep[AIA;][]{2012SoPh..275...17L},                 % Lemen et al. 2012 
which is part of the \textit{Solar Dynamics Observatory} 
\citep[\textit{SDO};][]{2012SoPh..275....3P}.        % Pesnell et al. 2012
AIA carries four telescopes and obtains full-Sun images at a 12 s cadence in seven different 
EUV filters and at a 24 s cadence in two UV filters. The spatial resolution is $\sim$1\farcs5
with a corresponding pixel size of 0\farcs6\,$\times$\,0\farcs6. We here in particular used 
data taken in the AIA 304\,\AA~(log\,$T$\,=\,4.7), 171\,\AA~(log\,$T$\,=\,5.8), and 94\,\AA~(log\,$T$\,=\,6.8) 
channels. The AIA data were downloaded from the Virtual Solar Observatory (VSO) in the level-1 
format, that is, they were already corrected for dark current and flat-fielded, de-spiked, and 
calibrated, but not exposure-time corrected. The filtergrams were then processed with aia$\_$prep.pro, 
which adjusts the images to a common plate scale so that they share the same centers and rotation 
angles. The AIA images are usually taken with fixed exposure times, which can cause saturation 
during the peak phase of the flare. To avoid this, the exposure times can be automatically
reduced during flares. We used these special short-exposure frames in our analysis when they 
were available. 

The \textit{Reuven Ramaty High Energy Solar Spectroscopic Imager} 
\citep[\textit{RHESSI};][]{2002SoPh..210....3L}        % Lin et al. 2002
has observed solar X-ray and gamma-ray emission from 3\,keV to 17\,MeV since its  
launch in 2002. It provides simultaneous imaging and spectroscopy
with high time-
and energy resolution as diagnostic tools of heated flare plasma with temperatures 
in excess of about 10\,MK and nonthermal bremsstrahlung that
is due to accelerated electrons.

We also used magnetograms from the Helioseismic and Magnetic Imager
\citep[HMI;][]{2012SoPh..275..207S}                    % Scherrer et al. 2012 
for context purposes. HMI data were also processed with aia$\_$prep.pro to place 
them on the same plate scale as the AIA images.

All data were precisely spatially co-aligned before further analysis. The coordinate 
system of AIA was used as reference. The position of the EIS slit relative to the AIA images was 
obtained using cross-correlation techniques. To verify the final alignment, space-time maps of selected EIS and 
AIA intensity channels were constructed using the derived slit positions and positions 
shifted for several pixels in all directions. These 2D maps were again cross-correlated to 
find the best match. This method allowed us to co-align EIS and AIA data with a precision 
of around 1$^{\prime\prime}$ , and the position of EIS slit center was found to be at 
$(x,y)\,=\,(448.6^{\prime\prime},-46.4^{\prime\prime})$. \textit{RHESSI} data are generally well aligned with 
the AIA filtergrams within 1$^{\prime\prime}$\,-\,2$^{\prime\prime}$.  

As the plasma parameters evolve quickly during the impulsive flare phase, it is important 
to state observation times precisely (especially if data from several instruments were 
taken with very different exposures). Therefore we always use the midpoint of the particular 
exposure or integration time (given in UT), respectively, if we refer to individual exposures 
of EIS, AIA, HMI, and \textit{RHESSI}. 
%__________________________________________________________________

\section{Results} \label{Results}
\subsection{Event overview} \label{Event overview}

\begin{figure}[!t]
   \centering
   \includegraphics[width=\hsize]{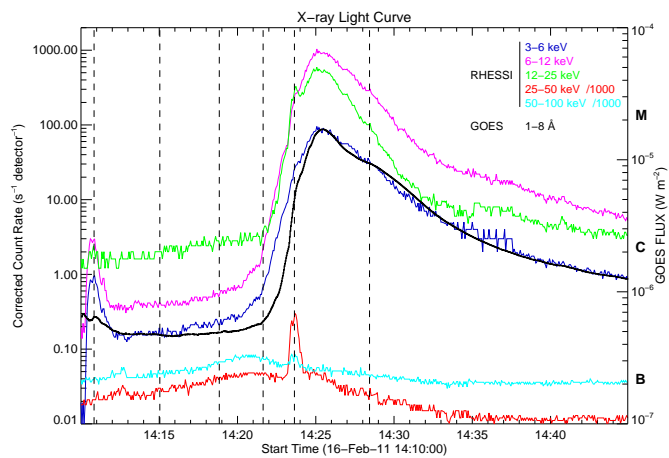}
      \caption{\textit{RHESSI} (3-100 keV, color-coded lines) and \textit{GOES} (1-8\,{\AA}, black line) 
               light curves showing the temporal evolution of the X-ray flux during 
               the M1.6 flare. The vertical dashed lines indicate the selected times for AIA 
               images and \textit{RHESSI} sources shown in Fig.\,\ref{aia_rhessi_from_young}.
              }
         \label{goes_rhessi_evol}
   \end{figure}

\begin{figure}[!t]
   \centering
   \includegraphics[width=\hsize]{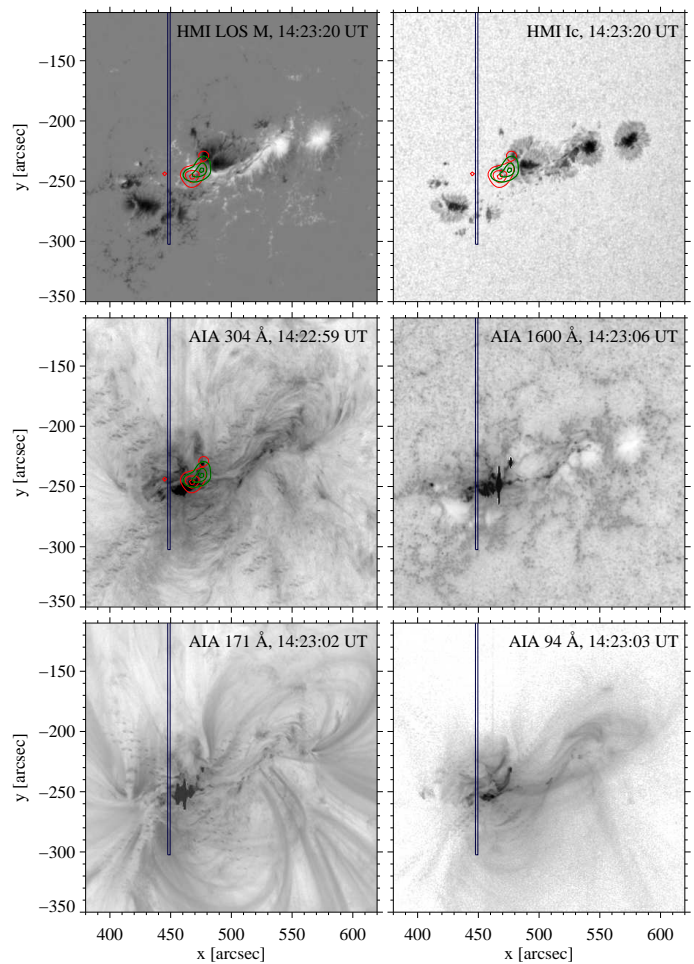}
      \caption{Overview images showing NOAA 11158 during the 
               main impulsive phase of the flare. 
               {\em Top left and right panels} show the HMI line-of-sight 
               magnetic field and continuum intensity, respectively. 
               {\em Middle and bottom panels} 
               are AIA filtergrams at different wavelengths (indicated in the 
               top right corner of each panel). They 
               are displayed on logarithmic scale and a reversed 
               intensity scaling. The times listed 
               in the upper right corner of each panel correspond 
               to the midpoint of the particular HMI or AIA exposure time.
               The overplotted contours show \textit{RHESSI} X-ray  
               images reconstructed in the 4-10\,keV (green) and 
               20-50\,keV (red) energy bands, with an integration 
               time of 40\,s, starting at 14:23:08\,UT. The contour levels 
               represent 10, 50, and 90\% of the peak intensity. The rectangle 
               outlines the lower part of the \textit{Hinode}/EIS slit. 
              }
         \label{rhessi_aia}
   \end{figure}

The temporal evolution of the \textit{RHESSI} 3--100\,keV count rates  
and \textit{GOES} 1-8\,\AA~SXR flux are plotted in Fig.\,\ref{goes_rhessi_evol}. The recorded 
\textit{GOES} 3-sec data show that the flare started at 14:19\,UT. The gradual increase in 
the flux changed to impulsive slightly before 14:22\,UT, and the flare peaked 
at 14:25\,UT (classifying the event as an M1.6 flare). Afterward, the X-ray 
flux decreased gradually. A similar temporal evolution is also
seen in 
the low-energy \textit{RHESSI} light curves (from 3 to 12\,keV), which are dominated by 
thermal emission from the hot coronal flare plasma. In contrast, 
the high-energy \textit{RHESSI} curves (25 to 100\,keV) show a different behavior. 
A gradual increase is followed by a sharp peak (especially the 25--50\,keV channel) at 
14:23:38\,UT. The \textit{RHESSI} high-energy light curves result from nonthermal 
bremsstrahlung emission from flare-accelerated electrons and are therefore a proxy for the evolution 
of the energy deposition rate in the flare. This implies that the main energy deposition occurred 
roughly 1.5\,min before the flare reached its soft X-ray (SXR) peak.   
This time delay may be attributed to the characteristic time of chromospheric evaporation flows, 
filling the flare loop (governed by the sound speed). From the \textit{RHESSI} 4--10\,keV images, 
derived over the flare HXR peak shown in Fig.\,\ref{rhessi_aia}, we estimate the half-length of 
the flare loop as about 10\,Mm. The fastest evaporation flows derived from the EIS spectroscopy 
(cf. Sect.\,\ref{EIS spectroscopy of the flare}) have speeds of 80 to 150\,km\,s$^{-1}$. The 
resulting loop filling time, derived from these numbers, lies in the range of 70 to 120\,s, 
which means that it is consistent with the observed delay between the peaks of the HXR and SXR curves. 
%
%______________________________________________ GOES+RHESSI 
%______________________________________________
%
%______________________________________________ RHESSI-AIA
%______________________________________________

Context images taken by the \textit{SDO}/HMI (magnetograms and continuum intensity)
and the \textit{SDO}/AIA (filtergrams at different wavelengths
and temperatures) 
instrument showing NOAA 11158 during the impulsive phase of the flare 
are shown in Fig.\,\ref{rhessi_aia}. The AIA filtergrams taken with 
short exposures (compared to standard observations) were used to visualize 
the impulsive phase of the flare. \textit{RHESSI} CLEAN images  
integrated over 40\,s starting at 14:23:08\,UT (i.e., covering the HXR peak)
are overplotted in Fig.\,\ref{rhessi_aia}. The flare was 
rather compact, showing signatures of two flaring footpoints (\textit{RHESSI} 
20-50\,keV contours) that are connected by a loop top source (\textit{RHESSI} 4-10\,keV 
contours).

\begin{figure*}[!t]
   \centering
   \includegraphics[width=17cm]{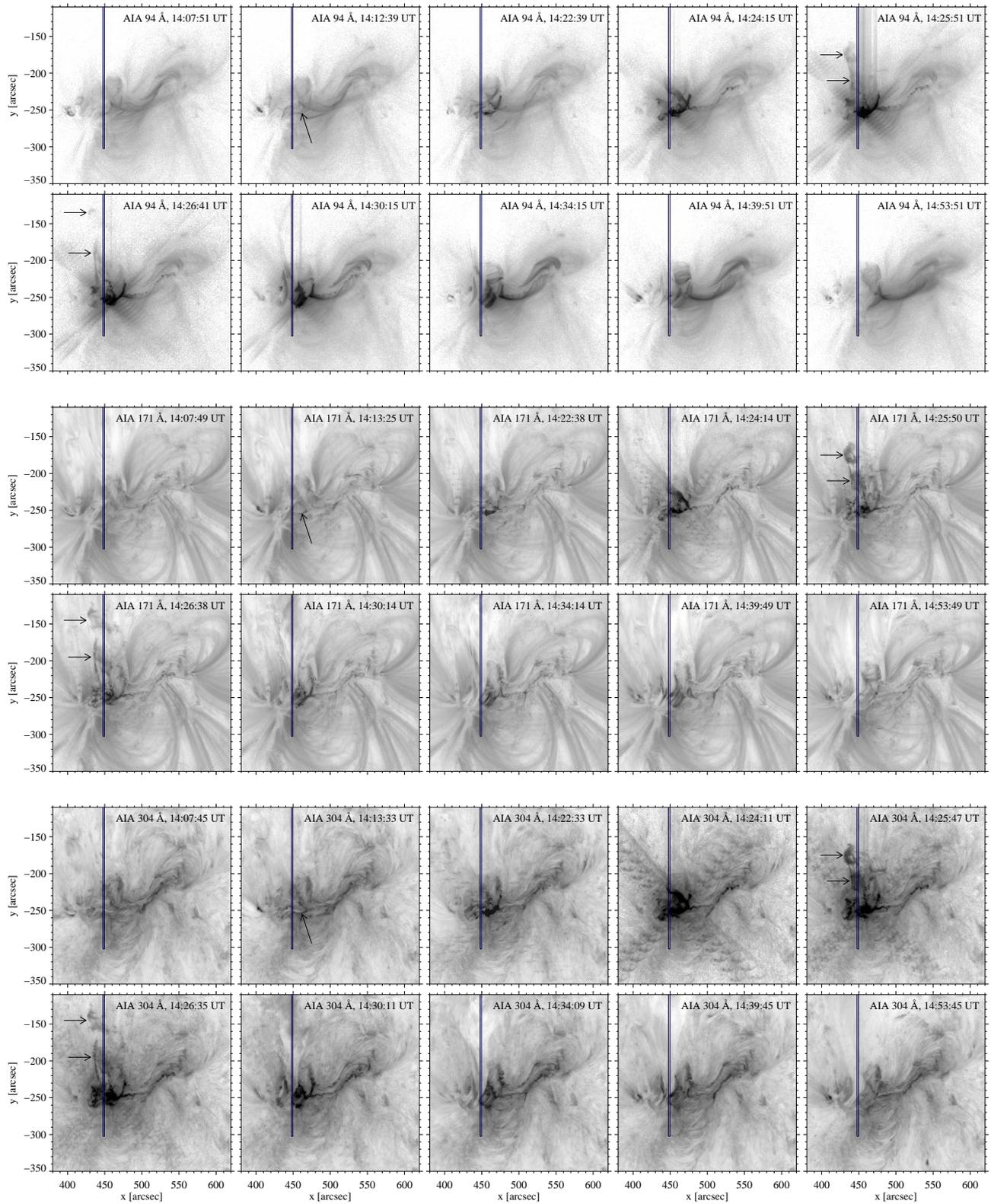}
      \caption{Sequence of AIA 94\,\AA~{\em (top two rows)}, 171\,\AA~{\em (middle two rows),} and 
               304\,\AA~{\em (bottom two rows)} filtergrams showing snapshots of the temporal 
               evolution of the flare. The midpoint of the recording time 
               of the particular images is presented in the upper right corner 
               of each panel. The arrow in the second panel of each 
               wavelength channel marks the appearance of a newly formed hot loop 
               structure. The arrows in the 5th and 6th panel point to 
               the erupting filament. The rectangles represent the position of 
               the EIS slit. An animation of the displayed wavelength channels is 
               shown in the attached movie.   
              }
         \label{aia_94_171_304_evolution}
   \end{figure*}

Figure \ref{aia_94_171_304_evolution} depicts the temporal evolution of the 
flare as observed in the AIA 94\,\AA, 171\,\AA, and 304\,\AA~filters. 
About 10 minutes before the flare, a newly formed loop-like structure appears in 
all these wavelength channels. 
The exact times of the first appearance of this feature are 14:09:57\,UT for 304\,\AA, 
14:10:13\,UT for 171\,\AA, and 14:10:51\,UT for 94\,\AA, and it is highlighted 
by an arrow in the second panel of the particular wavelength channels 
(Fig.\,\ref{aia_94_171_304_evolution}; note that filtergrams taken slightly later 
than the time of the first appearance of the loop-like structure are used because they allow 
an easier identification of the discussed structure). After its first occurrence, 
the brightness of the loop-like structure increased slowly until $\sim$14:22:30\,UT 
(the nominal flare onset). The actual flare site is co-spatial with 
the western footpoint of the loop-like structure and is located very close to the EIS slit 
position. As the flare evolved, filament material was ejected in the form of two clouds 
(marked by arrows in the 5th and 6th panels of each wavelength channel in 
Fig.\,\ref{aia_94_171_304_evolution}). The erupting filament material is 
traceable in the form of enhanced intensity 
in all analyzed channels and partly propagated along the EIS slit. 
Between $\sim$14:30\,UT and 14:40\,UT, the newly evolved system 
of the flare loops was partly covered by the EIS slit 
(best visible in the AIA 94\,\AA~channel).  

\begin{figure*}[!t]
   \centering
   \includegraphics[angle=90,width=\hsize]{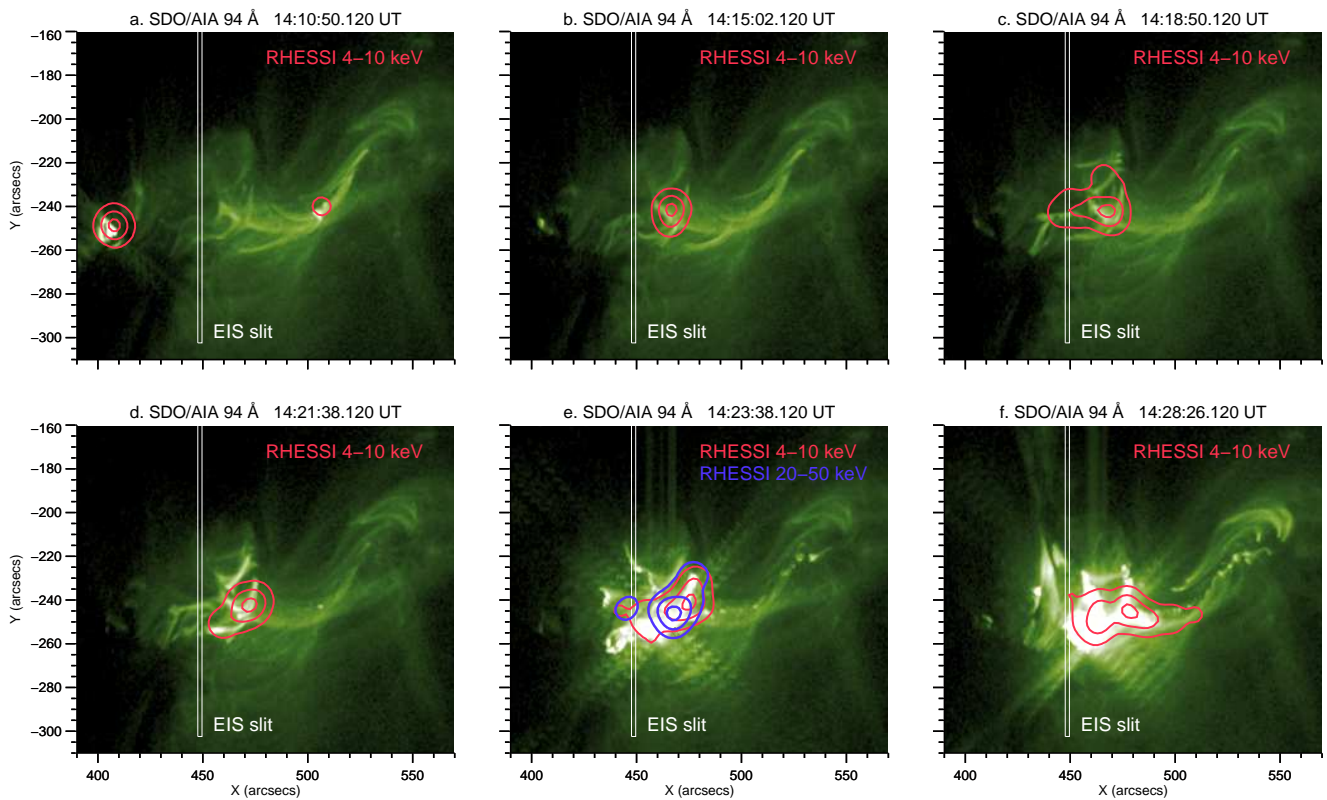}
      \caption{SDO/AIA 94\,{\AA} images at six different times, as indicated 
               in the X-ray light curves plotted in Fig.\,\ref{goes_rhessi_evol}. 
               These images show the evolution of flare 
               loops at a temperature around 6\,MK. \textit{RHESSI} X-ray sources 
               are shown as contours in red (4-10\,keV) and blue 
               (20-50\,keV). They show the locations of X-ray-emitting thermal plasma and 
               nonthermal electrons. The contour levels are 5, 30, 
               and 80\% of the peak intensity. The white rectangle depicts the lower part of 
               the \textit{Hinode}/EIS slit.
              }
         \label{aia_rhessi_from_young}
   \end{figure*}

The early flare evolution, covering the pre-flare and impulsive phase, is 
visualized in Fig.\,\ref{aia_rhessi_from_young}, showing a sequence of 
AIA\,94\,\AA~images together with co-temporal \textit{RHESSI} 4-10 and 20-50\,keV 
sources in the form of contours. The \textit{RHESSI} images were reconstructed using the Clean algorithm 
\citep{2002SoPh..210...61H}                         % Hurford et al. 2002
with detectors 2--7. The brightenings in the \textit{RHESSI} 4-10\,keV SXR image 
mark the flare onset at   
locations slightly away from the later main flare site (Fig.\,\ref{aia_rhessi_from_young}a). 
However,  the next panel (Fig.\,\ref{aia_rhessi_from_young}b) 
clearly shows a stable patch of 4-10\,keV emission exactly at the main flare site. The \textit{RHESSI} 
SXR source is situated very close to the position of the newly developed 
loop structure marked in Fig.\,\ref{aia_94_171_304_evolution}, which is later observed 
in AIA to be the commencement site of the main flare (cf. Fig.\,\ref{aia_rhessi_from_young},d).
%______________________________________________ A94, A171, A304 evolution
%______________________________________________

\subsection{EIS spectroscopy of the filament eruption} \label{EIS spectroscopy of the filament eruption}

\begin{figure*}[!ht]
   \centering
   \includegraphics[width=\hsize]{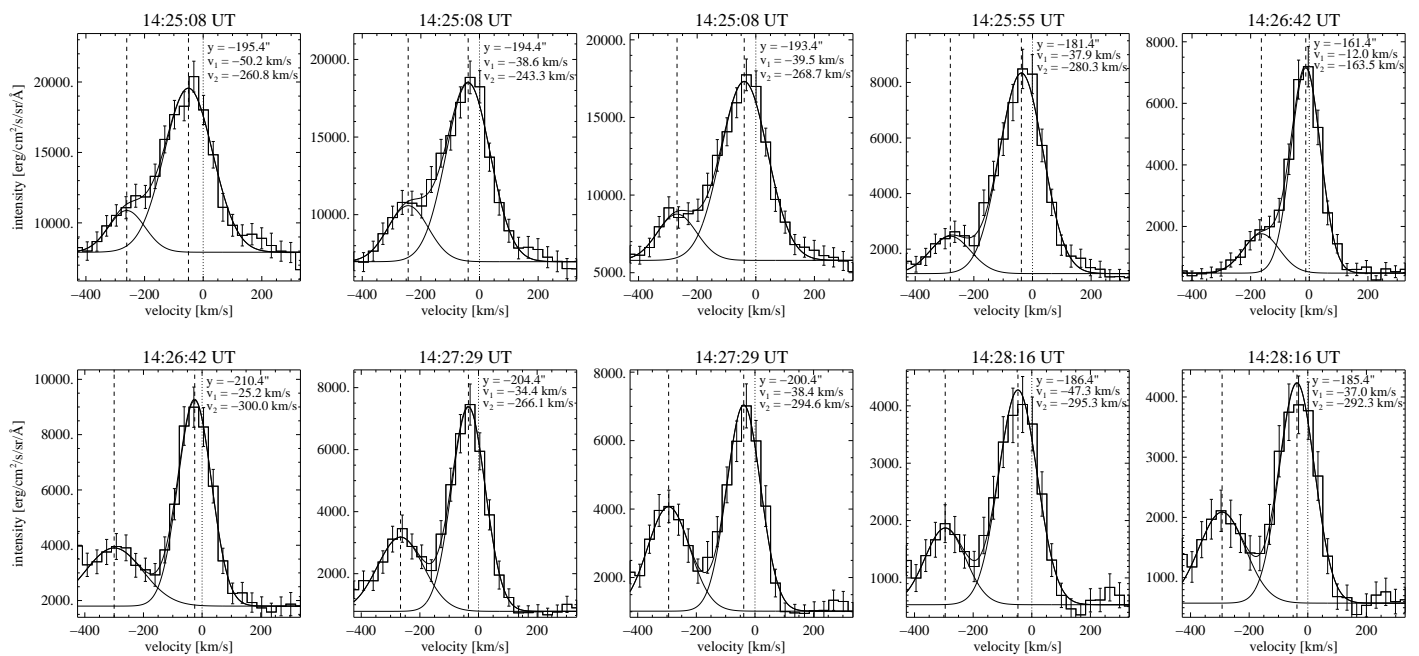}
      \caption{Representative examples demonstrating the two-component character of 
               the \ion{Fe}{xiii} 202\,\AA~spectral profiles related to the 
               erupting filament material ({\em the top row} represents 
               the leading and {\em the bottom row} the following part of the ejected 
               filament marked by the arrows in the 5th and 6th panel of
               Fig.\,\ref{aia_94_171_304_evolution}). The histograms show the observed 
               spectral profiles. The standard deviation of the line intensity is indicated by 
               error bars. The individual spectral components and the 
               resulting fits are plotted with solid lines. Vertical dotted lines 
               show the position of the zero Doppler shifts, while the vertical dashed 
               lines highlight the values of Doppler shifts determined for 
               particular components of the fit. The positions along 
               the slit together with the determined values of Doppler shifts 
               are given in the upper right corner of each panel.   
              }
         \label{profiles_ejecta}
   \end{figure*}

As mentioned before, the EIS observations captured part of the filament eruption. 
All \ion{Fe}{xiii} 202\,\AA~spectral line profiles in the corresponding parts of the 
EIS slit exhibit a two-component shape with the second Gaussian component to the 
spectral line profile shifted to shorter wavelengths. 
Examples of such spectral profiles are shown in Fig.\,\ref{profiles_ejecta}. The detected 
blueshifts correspond to average Doppler velocities of around $-250$\,km\,s$^{-1}$ and 
$-300$\,km\,s$^{-1}$ for the first and second cloud of ejected filament material, 
respectively. These values can be used as a lower estimate of the real speed 
of the filament eruption. We note that the \ion{Fe}{xvi} 262\,\AA~spectra  
were too noisy and weak in the filament to allow for a reliable analysis of this type. 
%______________________________________________ AIA 94 A plus RHESSI contours 
%______________________________________________

\subsection{EIS spectroscopy of the flare} \label{EIS spectroscopy of the flare}

\begin{figure*}[!t]
   \centering
   \includegraphics[width=\hsize]{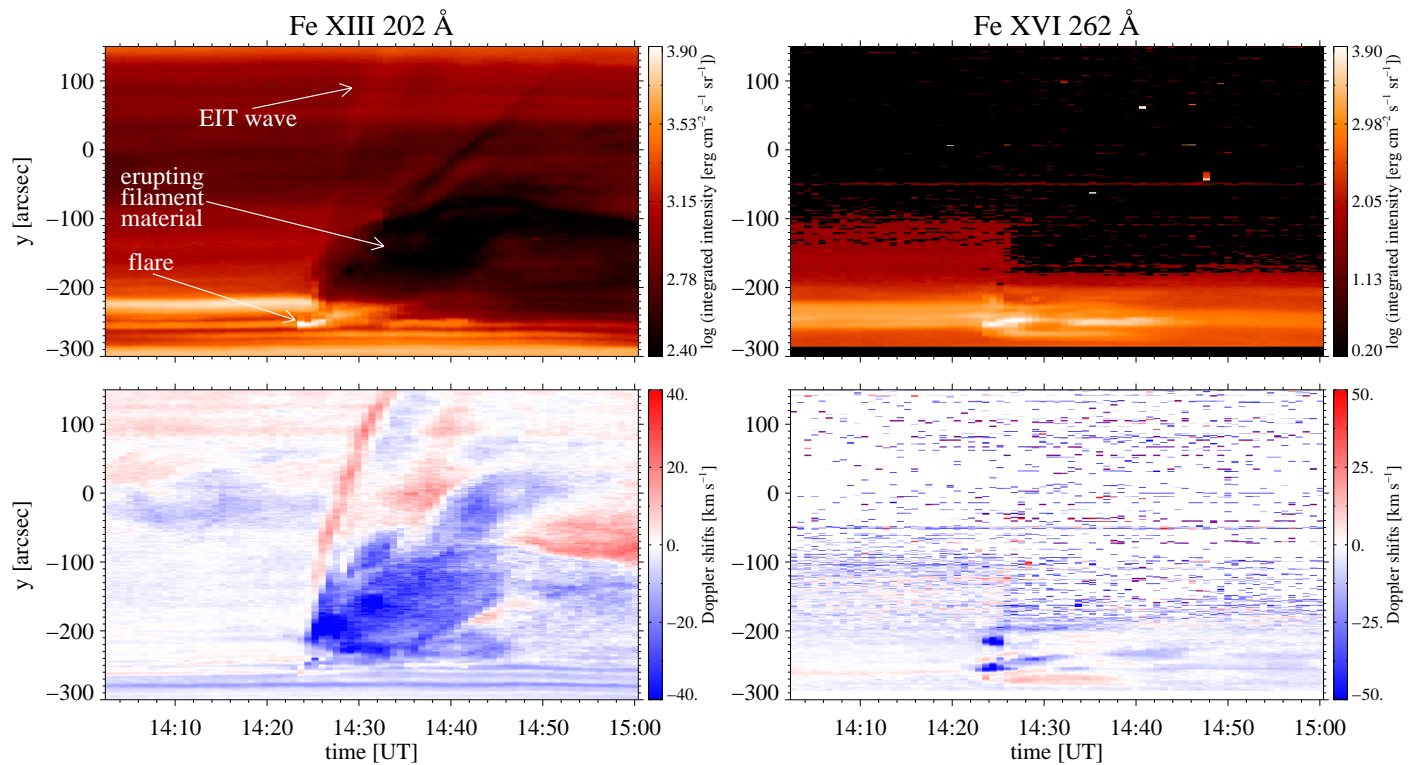}
      \caption{Time-space maps of the \ion{Fe}{xiii} 202\,\AA~and 
               \ion{Fe}{xvi} 262\,\AA~intensities {\em (top panels)} and 
               Doppler shifts {\em (bottom panels)}. The temporal evolution 
               of the observed flare is visible in the region between -260$^{\prime\prime}$ 
               and -250$^{\prime\prime}$ along the slit. Observed features (EIT wave, 
               erupting filament, and flare) are marked by arrows in the 
               \ion{Fe}{xiii} 202\,\AA~intensity map. 
              }
         \label{eis_2d_spectrograms}
   \end{figure*}

\begin{figure}[!ht]
   \centering
   \includegraphics[width=\hsize]{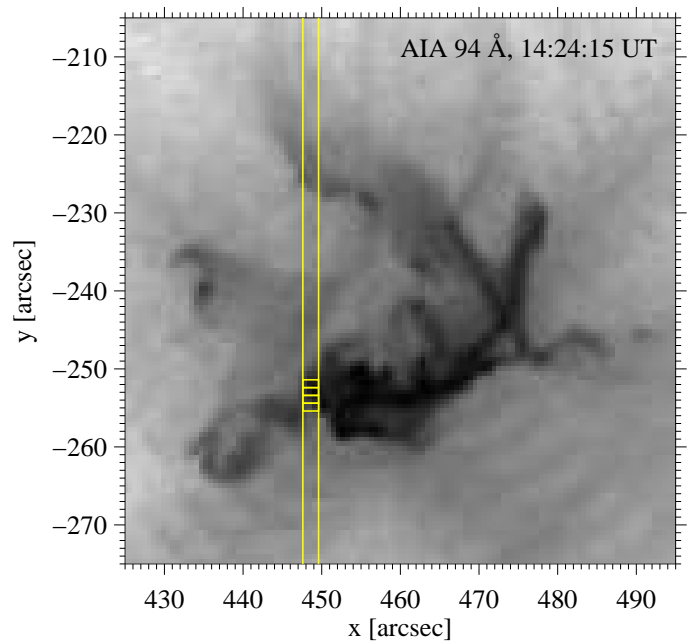}
      \caption{Position of the EIS slit (yellow vertical lines) superimposed on
               an AIA 94\,\AA~image, taken during the impulsive phase of the flare 
               at 14:24:15\,UT. The yellow rectangles highlight the EIS pixels we used to 
               study the temporal evolution of the intensities and Doppler shifts shown 
               in Fig.\,\ref{eis_spectral_param_evol}.  
              }
         \label{eis_aia_coalig}           
   \end{figure}

Figure\,\ref{eis_2d_spectrograms} gives an overview of the spectroscopic parameters 
(intensities and Doppler shifts), determined from the \ion{Fe}{xiii} 202\,\AA~and 
\ion{Fe}{xvi} 262\,\AA~spectral lines. The time-space maps 
show several structures such as the propagating EIT wave, the erupting filament 
material, and the flare site 
\citep[cf.][]{2011ApJ...743L..10V,                 % Veronig et al. 2011   
              2011ApJ...737L...4H}.                % Harra et al. 2011
In the following, we concentrate on a small fraction along the EIS slit 
(roughly between $-260^{\prime\prime}$ and $-250^{\prime\prime}$). Figure\,\ref{eis_aia_coalig} 
shows that this part of the EIS slit was co-spatial with one of the flare kernels 
(cf. also the attached movie). \textit{RHESSI} imaging also demonstrates that the eastern flare 
kernel (compare the 20-50\,keV contour) is partially located at the EIS slit 
(Fig.\,\ref{aia_rhessi_from_young}e).

\begin{figure*}[!t]
   \centering
   \includegraphics[width=16.2cm]{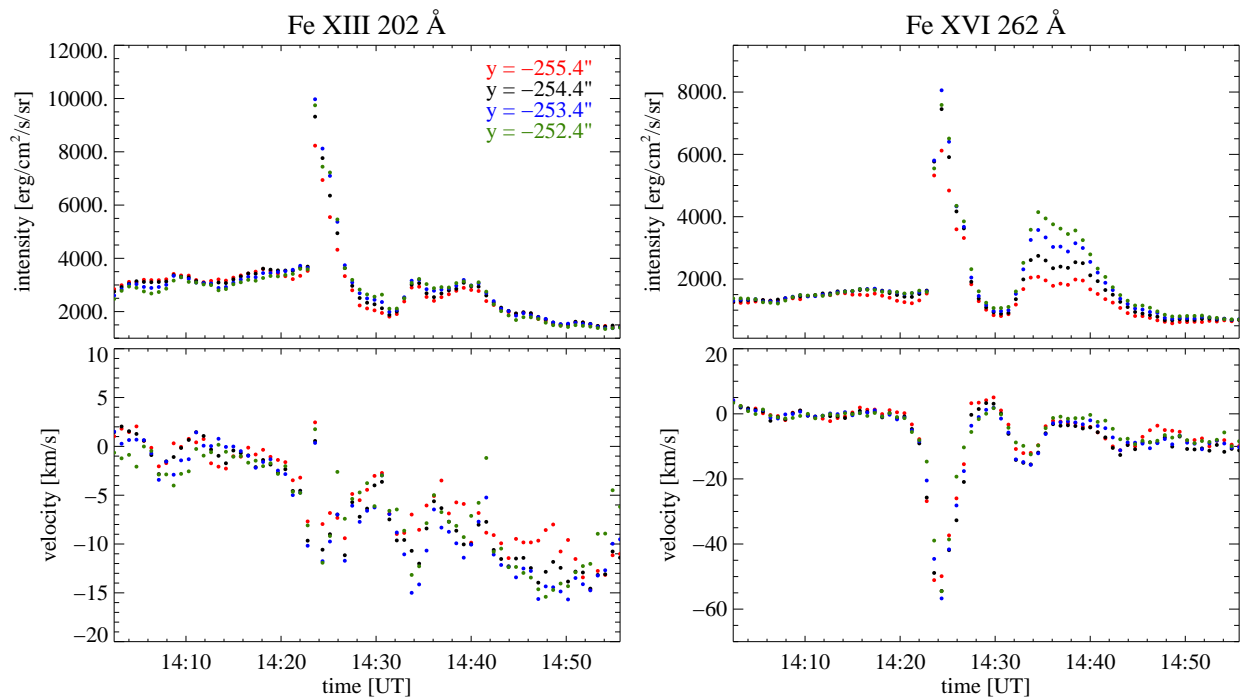}
      \caption{Temporal evolution of the \ion{Fe}{xiii} 202\,\AA~and 
               \ion{Fe}{xvi} 262\,\AA~intensities {\em (top panels)} and 
               Doppler shifts {\em (bottom panels)}, extracted from four 
               consecutive pixels along the EIS slit that cover the flaring 
               region. The pixels are indicated by yellow rectangles in 
               Fig.\,\ref{eis_aia_coalig}.
              }
         \label{eis_spectral_param_evol}
   \end{figure*}

Figure\,\ref{eis_spectral_param_evol} shows the temporal variations of the intensities 
{\em (top panels)} and Doppler shifts {\em (bottom panels)} that were extracted from four 
consecutive locations along the EIS slit covering the flare kernel. The variations   
represent the pre-flare, main impulsive phase, and post-flare time sequence. 
The exact location of the selected EIS pixels is marked in Fig.\,\ref{eis_aia_coalig}.  
%
%______________________________________________ Profiles_ejecta
%______________________________________________

The flare commencement in the selected pixels is clearly visible in Fig.\,\ref{eis_spectral_param_evol} 
in the form of a sharp increase in the intensities of the two lines we study. The intensity 
of the cooler line (\ion{Fe}{xiii} 202\,\AA) reaches its maximum at 14:23:34\,UT, the hotter 
line (\ion{Fe}{xvi} 262\,\AA) reaches it one exposure later 
(at 14:24:21\,UT). The Doppler shifts of the two lines behave differently. 
The \ion{Fe}{xiii} 202\,\AA~emission is subject to weak blueshifts 
during the pre-flare phase, which gradually increase from values close to 0\,km\,s$^{-1}$ 
at 14:14:12\,UT to velocities of around -10\,km\,s$^{-1}$, measured just before the peak of the 
spectral line intensity. We assume that this represents slowly expanding plasma that was 
heated before the impulsive phase of the flare. On the other hand, the \ion{Fe}{xvi} 262\,\AA~line 
exhibited no obvious Doppler shifts, indicating that the local atmospheric plasma at this 
temperature was still at rest during the pre-flare phase. Then, at the moment of the main intensity peak, 
the \ion{Fe}{xiii} 202\,\AA~line exhibits weak redshifts of 2-3\,km\,s$^{-1}$. 
These redshifts were only detected for a single exposure. Because the uncertainty in the 
Doppler shift calculation introduced by the HK method (i.e., wavelength drift compensation) 
corresponds to about 4.4\,km\,s$^{-1}$, however, the estimated redshifts are lower than the instrument 
resolution and are probably not reliable. Later on, again only blueshifts 
of about -10\,km\,s$^{-1}$ and lower (approaching almost zero velocity at $\sim$14:30\,UT) 
were present. In contrast, during the flare impulsive phase, the \ion{Fe}{xvi} 262\,\AA~velocities 
reveal only blueshifts. Their monotonic increase starts to appear roughly at 14:21\,UT, 
and they reached maximum values of $\sim$55\,km\,s$^{-1}$ at the moment of the intensity peak.
The velocities then started to decrease and revealed minimum values at the time interval 
14:27-14:30\,UT. 

\begin{figure*}[!t]
   \centering
   \includegraphics[width=\hsize]{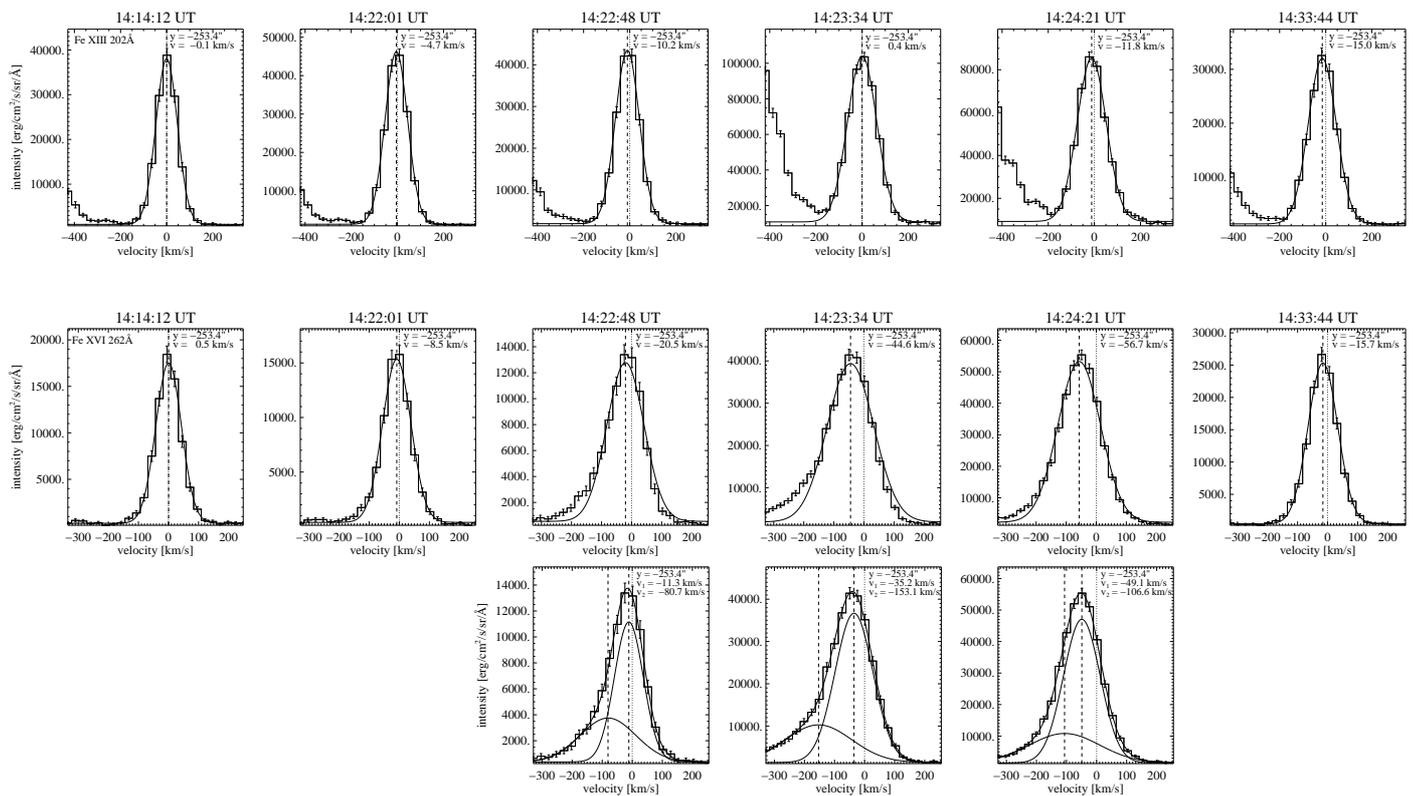}
      \caption{Representative examples of \ion{Fe}{xiii} 202\,\AA~{\em (top row)} 
               and \ion{Fe}{xvi} 262\,\AA~{\em (middle and bottom rows)} spectral 
               line profiles and their fits for particular phases 
               of the flare evolution, i.e., early pre-flare period {\em (first column)}, 
               impulsive phase {\em (second to fifth column)}, and decline phase 
               {\em (last column)}. The {\em top and middle rows} show the 
               profiles fitted by  a single-Gaussian function. The {\em bottom row} 
               displays results obtained from two-component fitting of the asymmetric 
               \ion{Fe}{xvi} 262\,\AA~spectral profiles detected during the impulsive 
               phase of the flare.  
               The histograms show the observed spectral profiles. The error bars mark  
               the standard deviation for each intensity value. Solid lines represent 
               final fits and individual spectral 
               components {\em (bottom row)}. Vertical dotted lines indicate
               the position of zero Doppler shifts, while the vertical dashed 
               lines highlight the values of Doppler shifts determined for 
               particular components of the fit. The positions along 
               the slit together with the determined values of Doppler shifts 
               are inserted in the upper right corner of each panel.   
              }
         \label{eis_spectra_during_flare}
   \end{figure*}

Figure\,\ref{eis_spectra_during_flare} shows representative spectra of each flare phase (pre-flare, 
impulsive, and decay). Close inspection reveals that the \ion{Fe}{xvi} 262\,\AA~spectral 
profiles exhibit a significant asymmetry during the impulsive phase. When approximated by two 
Gaussians, high-velocity components are derived, 
suggesting that a fraction of the plasma was moving upward with velocities of up to 80-150\,km\,s$^{-1}$. 
In contrast, only single-Gaussian spectral profiles were detected for the \ion{Fe}{xiii} 202\,\AA~line 
during the whole flare evolution (they were just slightly broadened during the impulsive phase). 
However, a significant intensity increase is visible in the very far blue wing of this line 
(Fig.\,\ref{eis_spectra_during_flare}, {\em top row}). If this spectral component were related  
to the flare, then it would represent upflows with extremely high Doppler shifts (more then 
450\,km\,s$^{-1}$), which are sometimes observed during flares in spectral lines formed at temperatures 
higher than 10\,MK, but not at temperatures sampled by \ion{Fe}{xiii} spectral lines. However, this 
enhancement might also be the signature of another spectral line that becomes strong when the local atmosphere began to be heated. 
The CHIANTI spectral line list suggests that it may be a
blend of \ion{Fe}{xi} and \ion{Fe}{xii} around 201.74\,\AA. This blend is at just the right wavelength 
to imply a blueshifted \ion{Fe}{xiii} component with a velocity corresponding to about 450\,km\,s$^{-1}$.
This explanation is also supported by the fact that the feature is visible 
well before the impulsive flare phase (e.g., at 14:14:12\,UT, as is shown in 
Fig\,\ref{eis_spectra_during_flare}). 

\begin{figure}[!th]
   \centering
   \includegraphics[width=\hsize]{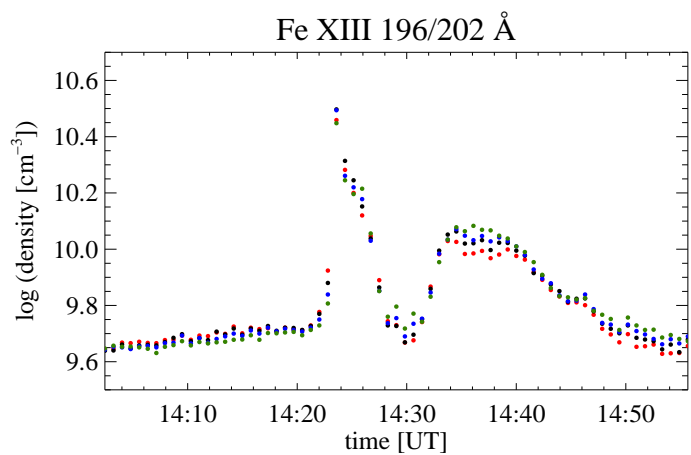}
      \caption{Temporal variations of the electron density determined using the 
               \ion{Fe}{xiii} 196/202\,\AA~line pair. The displayed values 
               were selected from the same spatial pixels as intensities 
               and Doppler shifts shown in Fig.\,\ref{eis_spectral_param_evol}. 
              }
         \label{eis_density_evol}
   \end{figure}

The impulsive phase of the flare evolution is accompanied by a rapid 
enhancement in the electron densities (Fig.\,\ref{eis_density_evol})  
computed from the intensity ratio of the \ion{Fe}{xiii} spectral lines 
observed at 196 and 202\,\AA. This line pair was selected for the density 
analysis instead of the available \ion{Fe}{xiii} 203/202\AA~line pair because it is more sensitive 
to the higher densities expected during the flare. The calculated electron densities rise 
from a pre-flare level of log(n$_{e}$)\,=\,$\sim$9.7 (i.e., 5.01$\times$10$^{9}$\,cm$^{-3}$) 
to values of up to log(n$_{e}$)\,=\,$\sim$10.5 (i.e. 3.16$\times$10$^{10}$\,cm$^{-3}$) within 
less than two\,minutes during the impulsive phase. 

During the decline phase, the Doppler shifts of both \ion{Fe}{xiii} 202\,\AA~and 
\ion{Fe}{xvi} 262\,\AA~lines exhibit secondary peaks of blueshifts with maximum values 
of $\sim$15\,km\,s$^{-1}$ in the time interval between 14:31\,UT and 14:36\,UT 
(Fig.\,\ref{eis_spectral_param_evol}). Spectral profiles typical for this phase 
are displayed in Fig.\,\ref{eis_spectra_during_flare} ({\em last column}). They 
confirm that the lines do not exhibit multi-component shapes at this time. The detected  
upflows correspond to a 
significant intensity increase in the \ion{Fe}{xvi} 262\,\AA~line. Gentle intensity 
growth is also visible in the \ion{Fe}{xiii} 202\,\AA~line. Moreover, the electron 
densities (Fig.\,\ref{eis_density_evol}) determined from the pair of \ion{Fe}{xiii} 
lines exhibit a significant 
peak with log(n$_{e}$)\,=\,$\sim$10.07 (i.e. 1.17$\times$10$^{10}$\,cm$^{-3}$). 
We note that the intensity peaks of both lines and the density peak are delayed, but they persist 
much longer than the corresponding peaks visible in the Doppler shifts (compare 
Figs.\,\ref{eis_spectral_param_evol} and \ref{eis_density_evol}). The 
increase of the electron densities and intensities together with the detected upflows 
can be explained by a process that fills the loops with expanding hot material that is due to 
chromospheric evaporation. This interpretation is also supported by the fact that a new 
system of loops was observed at the EIS slit location at times that correspond to 
the discussed secondary density enhancement (see also Fig.\,\ref{aia_94_171_304_evolution}). 
%
%______________________________________________ EIS 2d spectrograms, int, vel
%______________________________________________

\subsection{RHESSI X-ray spectroscopy of the flare} \label{RHESSI X-ray spectroscopy of the flare}

Figure\,\ref{rhessi_spectra_from_young} displays three \textit{RHESSI} X-ray spectra together 
with the fitting results to show the evolution of the hot plasma and accelerated 
electrons during the rising phase, around the HXR peak time, and during the decay phase. These spectra were derived from detector 4 with 20\,s integration time and were
fit with an isothermal component and an nonthermal thick-target model
\citep{2003ApJ...595L..97H}.                        % Holman et al. 2003
%
%______________________________________________ coalig. 
%______________________________________________
A second isothermal component was added to improve the fitting result whenever 
necessary (Fig.\,\ref{rhessi_spectra_from_young}c). Following 
\cite{2013ApJ...765...37F}                          % Feng et al. 2013
and 
\cite{2012ApJ...759...71E},                         % Emslie et al. 2012
we estimated the nonthermal energy flux based on the fitting result around the HXR peak, which 
gives a value of 7.71$\times$10$^{27}$\,erg\,s$^{-1}$.
Taking the area enclosed by the 50\% contour in the 20-50\,keV image (Fig.\,\ref{aia_rhessi_from_young}e) as the 
cross section of the flaring loops, we estimated a nonthermal energy flux density of 
1.34$\times$10$^{10}$\,erg\,s$^{-1}$\,cm$^{-2}$, which is close to 
the threshold of 10$^{10}$\,erg\,s$^{-1}$\,cm$^{-2}$ between explosive and gentle evaporation 
derived from the hydrodynamic simulations by 
\cite{1985ApJ...289..414F}.                          % Fisher er al. 1985
%__________________________________________________________________

\section{Discussion} \label{Discussion}

The spectroscopic observations of the M1.6 flare under study revealed 
that the intensity maxima of the two analyzed spectral lines with different 
formation temperatures were not reached simultaneously (the intensity maximum 
of the hotter line follows the peak of the colder line by one exposure, i.e., by about 50\,s). 
However, both these times correspond well with the HXR peak time at 
14:23:38\,UT (\ion{Fe}{xiii} 202\,\AA~intensity peaks at 14:23:34\,UT
and  \ion{Fe}{xvi} 262\,\AA~intensity at 14:24:21\,UT). This suggests that the 
part of the EIS slit covering the flare ribbon (see Fig.\,\ref{eis_aia_coalig}) 
is located close to the site of the strongest energy deposition. 

We detected strong upflows of up to $\sim$55\,km\,s$^{-1}$ in the hotter line (\ion{Fe}{xvi} 262\,\AA) 
and no significant Doppler shifts (estimated downflows of 2-3\,km\,s$^{-1}$ are 
below the resolution of the spectrograph and thus are not very reliable) in the cooler line 
(\ion{Fe}{xiii} 202\,\AA) during the impulsive phase of the flare. Moreover, a high -speed component was discovered in the \ion{Fe}{xvi} 262\,\AA~spectra, 
suggesting that the flaring material expanded with velocities of up to 
80-150\,km\,s$^{-1}$. In contrast, only single-component spectra were detected 
in the \ion{Fe}{xiii} 202\,\AA~line during the impulsive and the decay 
phase of the flare evolution. 

A dependency of the Doppler velocity directions on the formation temperature 
of spectral lines has been observed during the impulsive phases of flares.  
\cite{2005ApJ...625.1027K}   % Kamio et al. (2005)
reported that they did not detect any significant changes in the Doppler shifts 
of the \ion{Mg}{ix} spectral line. Therefore, they suggested that the observed 
plasma at log\,$T$\,=\,6.0 was close to the intermediate temperature between the upflowing 
hot plasma (log\,$T$\,=\,7.0) and the downflowing chromospheric and transition region plasma 
(log\,$T$\,=\,4.0-5.0). 
\cite{2006ApJ...638L.117M}   % Milligan et al. (2006)
found strong blueshifts of up to $\sim$250\,km\,s$^{-1}$ co-spatial with flare ribbons using  
the \ion{Fe}{xix} spectral line (forming at log\,$T$\,=\,6.9) and 
weak redshifts for the \ion{He}{i} and \ion{O}{v} spectral lines (sensitive to 
chromospheric and transition region temperatures, respectively). This is again suggestive of the temperature regime where Doppler velocities change from downflows to 
upflows. This was also confirmed by the first study of the impulsive phase of a flare 
observed with the EIS spectrometer.    
\cite{2009ApJ...699..968M} % Milligan & Dennis (2009)
observed footpoints of a C1.1 flare and found that the Doppler shifts revealed a change from 
redshifts to blueshifts at around log\,$T$\,=\,6.3. In particular, they showed that the emission from spectral 
lines formed in the temperature range log\,$T$\,=\,6.3-7.2 exhibited clear blueshifts that 
scaled with temperature and reached velocities of $>$250\,km\,s$^{-1}$ for the hottest spectral
line \ion{Fe}{xxiv}. On the other hand, the emission formed at temperatures log\,$T$\,=\,4.7-6.2 
was found to be redshifted. 
%
%______________________________________________ EIS evol. of spectral char (int, vel)
%______________________________________________
In contrast, a detailed study of an M1.1 flare performed by 
\cite{2013ApJ...766..127Y}   % Young et al. (2013)
suggested that the change from redshift to blueshift appears between the \ion{Fe}{x} and \ion{Fe}{xii} 
spectral lines, that is, at a temperature of about log\,$T$\,=\,6.1 (see Fig.\,7 in their work). 
The lower temperature of the transition from redshifts to blueshifts was also reported by 
\cite{2011A&A...532A..27G}.   % Graham et al. (2011)
They observed small downflows at temperatures below \ion{Fe}{xiii} (log\,$T$\,=\,6.2) and upflows 
of up to $\sim$140\,km\,s$^{-1}$ at higher temperatures. Our results indicate that the intermediate 
temperature separating hot upflowing and cooler downflowing material is approximately at log\,$T$\,=\,6.2 
because we did not detect any significant flows in the \ion{Fe}{xiii} spectral line. 

The observed change of the Doppler velocity direction during the impulsive phase of the flare 
is interpreted within the model of explosive chromospheric evaporation. In this scenario, 
hotter plasma rises toward the corona and cooler plasma falls toward the photosphere to preserve momentum balance.   

Our analysis of the \textit{RHESSI} spectra and images suggests that the energy flux deposited 
by the beam of accelerated electrons to the lower atmospheric layers was 
1.34$\times$10$^{10}$\,erg\,s$^{-1}$\,cm$^{-2}$ during the flare HXR peak. This value is very 
close to the theoretical threshold of 10$^{10}$\,erg\,s$^{-1}$\,cm$^{-2}$ between gentle and 
explosive evaporation determined by 
\cite{1985ApJ...289..414F}.               % Fisher at al. 1985 
It is therefore somewhat puzzling that we found clear signatures of explosive 
evaporation under these conditions. 
%______________________________________________ EIS spectra during flare
%______________________________________________

\begin{figure*}[!th]
   \centering
   \includegraphics[angle=90,width=\hsize]{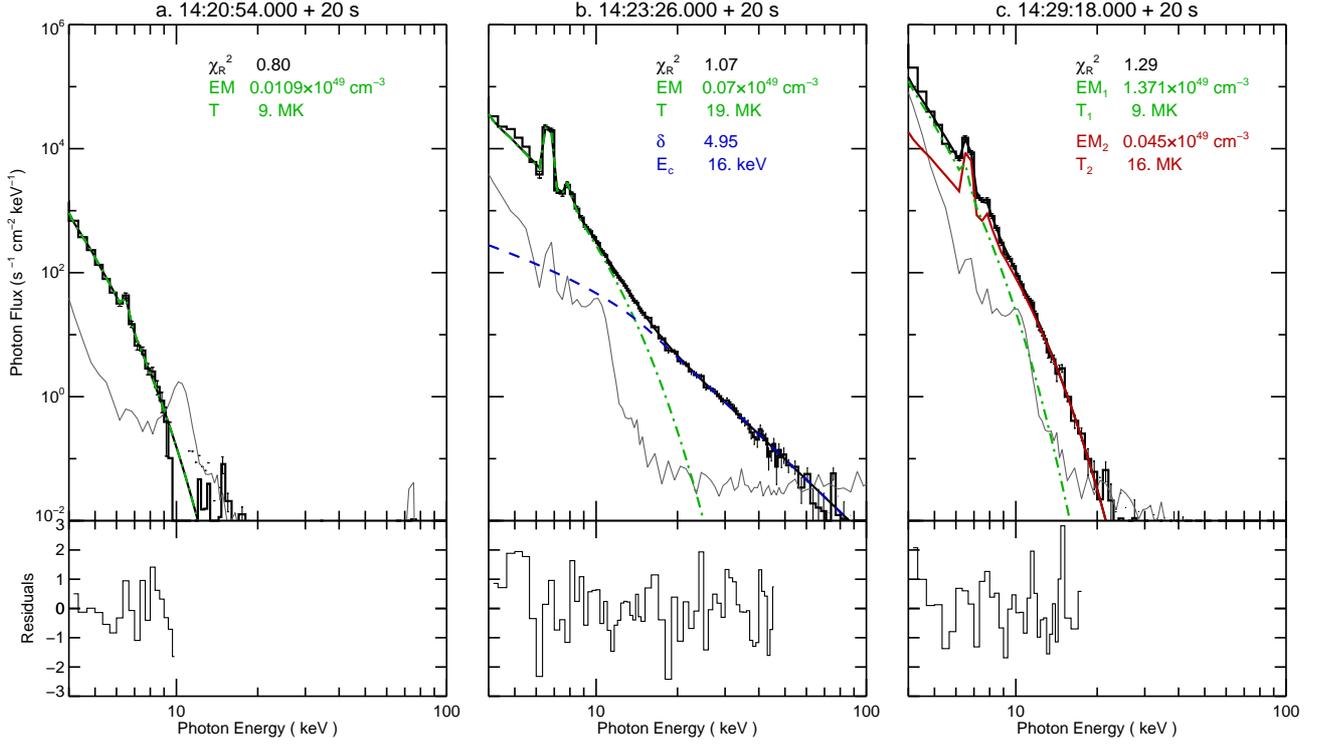}
      \caption{{\em Panel a to c, top:} Reconstructed X-ray spectra and fitting results. 
               The three examples shown here were selected at the rising phase, the HXR 
               peak, and the decay phase. These spectra were derived from detector 
               4 with 20\,s integration time (histogram) and fitted with an isothermal 
               component (green, dot-dashed line) and a nonthermal thick-target model 
               (blue, dashed line). A second isothermal component (purple, solid line) 
               was added when necessary for a better fit. The gray curves represent the background. 
               The obtained parameters, such as reduced chi-square, plasma temperature $T$, 
               emission measure $EM$, power-law distribution index $\delta$ of the nonthermal 
               electrons, and low-energy cutoff $E_c$, are listed in the top right corner of each 
               panel. {\em Panel a to c, bottom:} Residuals from the fitting.
              }
         \label{rhessi_spectra_from_young}
   \end{figure*}

However, the results of 
\cite{1985ApJ...289..414F}                % Fisher at al. 1985
were derived under several assumptions. A critical one is the 
usage of a fixed low-energy cut-off (20\,keV) in their simulation. This assumes an abundance of high-energy electrons in the beam, 
but no particles with energies below 20\,keV. In a recent study, 
\cite{2015ApJ...808..177R}                 % Reep et al. (2015) 
performed detailed hydrodynamical simulations to examine the atmospheric response 
to heating by different isoenergetic beams of accelerated electrons. In particular, 
they analyzed the role of electron energy and stopping depths in the two regimes of 
chromospheric evaporation. They found that the threshold between explosive and 
gentle evaporation found by 
\cite{1985ApJ...289..414F}                 % Fisher at al. 1985
depends quadratically on the electron energy and linearly on the beam flux. 
The threshold of $\sim$10$^{10}$\,erg\,s$^{-1}$\,cm$^{-2}$ was confirmed for beams 
formed by electrons with energy 20\,keV. But lower energy fluxes are sufficient to drive 
explosive evaporation if electrons with energies below 20\,keV are considered. 
We recall that in the flare under study, the fits to the \textit{RHESSI} spectra 
indicate a low-energy cut-off of 16\,keV (cf. Fig.\,\ref{rhessi_spectra_from_young}b), 
which is an upper limit, and the real cut-off might be even lower.
%______________________________________________ EIS density evolution 
%______________________________________________ 

Moreover, 
\cite{2015ApJ...808..177R}                 % Reep et al. (2015) 
pointed out that lower energy electrons are more efficient in heating flaring 
loops and driving dense plasma into the corona. For example, if electron beams with 
a constant energy flux density of 10$^{10}$\,erg\,s$^{-1}$\,cm$^{-2}$ were considered, 
then those simulations using a high-energy electron beam showed only small electron 
density enhancements during the flare. In contrast, significant density increase only appears 
during a flare for beams with electron energies of 15\,keV and below. This agrees 
with our result, which shows a clear density enhancement co-temporal with the HXR peak, 
that is, for the time when a low-energy cut-off of $\lesssim$16\,keV was estimated from the 
\textit{RHESSI} spectra. 

The higher efficiency of the low-energy electrons in triggering explosive 
evaporation was explained in the following way by 
\cite{2015ApJ...808..177R}.                % Reep et al. (2015)
Electrons of lower energy are 
effectively stopped higher up in the atmosphere and deposit their energy 
in the low-density plasma which has less inertia and has a lower heat capacity. 
Consequently, the plasma pressure starts to rise and explosive evaporation 
sets in. On the other hand, the stopping depth of high-energy electrons can 
extend deep into the solar atmosphere where the ambient density is much higher. 
In addition, this part of the atmosphere has a larger heat capacity and stronger 
radiative losses. Thus, the deposited energy is substantially lower than the local thermal energy. This explains why we observed clear 
evidence of explosive evaporation although the estimated energy flux density was 
close to 10$^{10}$\,erg\,s$^{-1}$\,cm$^{-2}$. 

We found significant enhancement of the electron densities during the impulsive flare phase, 
increasing from the pre-flare value of 
$\sim$5.01$^{+0.140}_{-0.137}\times$10$^{9}$\,cm$^{-3}$ to 
$\sim$3.16$^{+0.141}_{-0.135}\times$10$^{10}$\,cm$^{-3}$ within less than two\,minutes. This maximum 
occurs at the same time as the \ion{Fe}{xiii} 202\,\AA~velocities exhibit 
weak redshifts and the intensities peak sharply. However, it appears about 50\,s (i.e., one exposure) 
earlier than the \ion{Fe}{xvi} 262\,\AA~intensities and Doppler shifts reached their maxima.
%______________________________________________ RHESSI spectra 
%______________________________________________

The densities were determined using the 
\ion{Fe}{xiii} 196/202\AA~line pair and therefore correspond to temperatures around 1.58\,MK
(log\,$T$\,=\,6.2). Similar values of electron densities were determined by 
\cite{2011A&A...532A..27G},   % Graham et al. (2011)
who measured density enhancements of up to a few 10$^{10}$\,cm$^{-3}$ in the flare footpoints at 
temperatures of $\sim$1.5-2\,MK. 
\cite{2010ApJ...719..213W}   % Watanabe et al. (2010)
also observed an increase of the electron densities of up to 3$\times$10$^{10}$\,cm$^{-3}$ at 
evaporation sites.                   
\cite{2013ApJ...766..127Y}   % Young et al. (2013) 
used the \ion{Fe}{xiv} 264.79/274.20\,\AA~(formed at 2\,MK) spectral line ratio 
and determined electron densities of $\sim$3.4$\times$10$^{10}$\,cm$^{-3}$ 
co-spatial with a flare kernel. We note that while the findings of the latter authors were based on  
raster observations, we used sit-and-stare measurements for our study. 
The first EIS measurement of a flare using the sit-and-stare observing mode was performed by 
\cite{2013ApJ...762..133B}.    % Brosius (2013) APJ 762
He observed a C1 flare and determined the electron densities 
using a pair of \ion{Fe}{xiv} spectral lines. The analysis yielded an average pre-flare value 
of 3.75$\times$10$^{9}$\,cm$^{-3}$ and an average flare value of 4.94$\times$10$^{9}$\,cm$^{-3}$.
The reported lower density increase is very probably related to the fact that 
the flare under study was a \textit{GOES} C-class event.  On the other hand, analysis of 
another C1 flare presented by 
\cite{2013ApJ...777..135B}     %Brosius (2013) APJ 777  
showed that the electron densities derived again from a pair of \ion{Fe}{xiv} spectral lines
increased by an order of magnitude from the pre-flare value of 3.43$\times$10$^{9}$\,cm$^{-3}$
to its maximum impulsive phase value of 3.04$\times$10$^{10}$\,cm$^{-3}$ in two minutes. 
This is in quantitative agreement with our findings, although we analyzed a stronger 
(\textit{GOES} M1.6) flare. Therefore it suggests that the flare-induced changes of the electron 
densities are not strongly dependent on the flare energy.

The observed enhancement of the electron density during the main phase of the 
flare can be explained in the following way. Thermal and nonthermal energy released 
during the impulsive phase of the flare is deposited at deeper layers of the solar 
atmosphere, which consequently heat up. Thus, emission at 
coronal temperatures is released from much denser regions, explaining the observed 
increase. An alternative explanation is that the high-energy electrons that  
produce the observed HXR emission during the flare impulsive phase can easily 
penetrate the partially neutral atmosphere below where they are effectively stopped by 
collisions. This process can contribute to the ionization of the surrounding plasma 
and also lead to local heating and can be partly responsible for the measured 
enhancements of electron densities. 
   
After the impulsive flare phase, we found peaks of blueshifts reaching $\sim$15\,km\,s$^{-1}$ 
for the two spectral lines under study at the time interval between 14:31\,UT and 14:36\,UT. 
These were followed by intensity enhancements in both lines that appear two exposures 
(i.e., $\sim$100\,sec) later and by a significant increase of the electron densities, 
with the same time delay. We found that the footpoints of the flare loops were 
located co-spatial with the EIS slit at that particular moment. Therefore these observational facts 
provide evidence that the heated flaring plasma was upflowing and thus filling 
newly created magnetic loops that subsequently emit the observed radiation. 
%__________________________________________________________________

\section{Conclusions} \label{Conclusions}

We presented EIS sit-and-stare spectroscopy of an M1.6 flare together with 
\textit{RHESSI} X-ray observations. During the impulsive phase, we detected 
different behavior of Doppler shifts in the spectral lines under study, which are 
formed at slightly different temperatures. In particular, we found insignificant 
downflows in the \ion{Fe}{xiii} 202.044\,\AA~(log\,$T$\,=\,6.2) line and 
strong upflows in the \ion{Fe}{xvi} 262.980\,\AA~(log\,$T$\,=\,6.4) line. These flows were related 
to strong intensity increases observed in both lines and also to a significant increase 
of the electron densities measured at 1.56\,MK from about 5.01$\times$10$^{9}$\,cm$^{-3}$ to 
3.16$\times$10$^{10}$\,cm$^{-3}$. In summary, the spectroscopic analysis suggests that 
explosive chromospheric evaporation took place during the flare peak. However, \textit{RHESSI}
X-ray spectroscopy interpreted with a collisional thick target model provided an estimate 
of the nonthermal energy flux density on the order of $\sim$10$^{10}$\,erg\,s$^{-1}$\,cm$^{-2}$  
, meaning that the energy deposited in the lower solar atmosphere was on the level considered as the 
threshold between gentle and explosive evaporation in hydrodynamic simulations
\citep[e.g.,][]{1985ApJ...289..414F}.         % Fisher at al. 1985
These findings provide evidence that the response of the flaring atmosphere strongly 
depends on the properties of the heating electron beams, which agrees with the predictions 
derived from recent numerical simulations 
\citep{2015ApJ...808..177R}.                 % Reep et al. (2015)  
During the decline phase, we detected a secondary peak of intensities and electron densities 
preceded by upflows in both observed lines. These were interpreted as signatures of flare 
loops filled by the hot material that expands as a result of chromospheric evaporation.  
%__________________________________________________________________

\begin{acknowledgements}
      This work was supported by the project of the \"{O}sterreichischer
      Austauschdienst (OeAD) and the Slovak Research and Development
      Agency (SRDA) under grant Nos. SK 16/2013 and SK-AT-0003-12. 
      P.G. acknowledges the support from grant VEGA 2/0004/16 of the
      Science Grant Agency and from the project of the Slovak Research 
      and Development Agency under the Contract No. APVV-0816-11.
      A.M.V. and Y.S. gratefully acknowledge support from the Austrian 
      Science Fund (FWF) P27292. Y.S. also acknowledges the Thousand 
      Young Talents Plan, a sub-program of the “Recruitment Program of Global 
      Experts” (1000 Talent Plan), and 11233008 from NNSFC.
      J.K.T. acknowledges support from Austrian 
      Science Fund (FWF) P25383-N27. This article was created by the 
      realization of the project ITMS No. 26220120009, based on the supporting 
      operational Research and development program financed from the European 
      Regional Development Fund. The authors thank an anonymous 
      referee for constructive comments and valuable suggestions that improved this paper.
      \emph{Hinode} is a Japanese mission developed and launched by ISAS/JAXA, 
      with NAOJ as domestic partner and NASA and STFC (UK) as international 
      partners. It is operated by these agencies in co-operation with ESA and
      NSC (Norway).
      \emph{SDO} is a mission for NASA’s Living With a Star (LWS) Program. 
      CHIANTI is a collaborative project involving the NRL (USA), the Universities 
      of Florence (Italy) and Cambridge (UK), and George Mason University (USA). 
      \emph{RHESSI} is a NASA small explorer mission.
      This research has made use of NASA’s Astrophysics Data System.
\end{acknowledgements}

%__________________________________________________________________

\bibliography{aa27403-15}

%__________________________________________________________________

\end{document}